\documentclass[preprintnumbers,amsmath,amssymbm,prd]{revtex4}
\usepackage{epsfig}
\usepackage{graphicx}
\usepackage{amssymb}

\begin{document}
\title{Spin-charge induced scalarization of Kerr-Newman black-hole spacetimes}
\author{Shahar Hod}
\affiliation{The Ruppin Academic Center, Emeq Hefer 40250, Israel}
\affiliation{ }
\affiliation{The Hadassah Institute, Jerusalem 91010, Israel}
\date{\today}

\begin{abstract}
\ \ \ It has recently been demonstrated that Reissner-Nordstr\"om black holes 
in composed Einstein-Maxwell-scalar field theories can support static scalar field configurations 
with a non-minimal negative coupling to the Maxwell electromagnetic invariant 
of the charged spacetime. 
We here reveal the physically interesting fact that scalar field configurations 
with a non-minimal {\it positive} coupling to the spatially-dependent Maxwell electromagnetic invariant ${\cal F}\equiv F_{\mu\nu}F^{\mu\nu}$
can also be supported in black-hole spacetimes. Intriguingly, it is explicitly proved that the positive-coupling 
black-hole spontaneous scalarization phenomenon is induced by a non-zero combination $a\cdot Q\neq0$ 
of {\it both} the spin $a\equiv J/M$ and the electric charge $Q$ of the central supporting black hole. 
Using analytical techniques we prove that 
the regime of existence of the positive-coupling spontaneous scalarization phenomenon of 
Kerr-Newman black holes with horizon radius $r_+(M,a,Q)$ and a non-zero electric charge $Q$ (which, in principle, 
may be arbitrarily small) is determined by the {\it critical onset line} $(a/r_+)_{\text{critical}}=\sqrt{2}-1$. 
In particular, spinning and charged Kerr-Newman black holes in the composed Einstein-Maxwell-scalar field theory 
are spontaneously scalarized by the positively coupled fields in the dimensionless charge regime $0<{{Q}\over{M}}\leq\sqrt{2\sqrt{2}-2}$ 
if their dimensionless spin parameters lie above 
the critical onset line ${{a(Q)}\over{M}}\geq \big[{{a(Q)}\over{M}}\big]_{\text{critical}}={{1+\sqrt{1-2(2-\sqrt{2}){(Q/M)}^2}}\over{2\sqrt{2}}}$.
\end{abstract}
\bigskip
\maketitle

\section{Introduction}

It is well known, based on the mathematically elegant no-hair theorems presented in
\cite{Bek1,Her1,Hodstationary,BekMay,Hod1}, that asymptotically flat black-hole spacetimes in 
the composed Einstein-Maxwell-scalar field theory 
cannot support spatially regular static configurations of minimally coupled massless scalar fields. 
The most general black-hole solution of the non-linearly coupled Einstein-Maxwell--massless-scalar field equations is therefore 
described by the bald (scalarless) Kerr-Newman black-hole spacetime \cite{ThWe,Chan}.

Intriguingly, it has recently been revealed in the physically important works \cite{Herch,Herr} (see also \cite{aaa1,aaa2,Hodsc1,Hodsc2,Hodnrx}) 
that spherically symmetric charged black holes in composed Einstein-Maxwell-scalar field theories 
whose actions contain a direct non-minimal coupling term $f(\phi){\cal F}$ [see Eq. (\ref{Eq1}) below] 
between the scalar field $\phi$ and the Maxwell electromagnetic invariant ${\cal F}$ can support spatially regular configurations of the 
non-minimally coupled scalar fields. 

In particular, it has been established \cite{Herch,Herr,aaa1,aaa2,Hodsc1,Hodsc2,Hodnrx} that the critical boundary 
between the familiar (scalarless) black-hole solutions of the Einstein-Maxwell field equations and the hairy (scalarized) 
black-hole spacetimes that characterize the composed Einstein-Maxwell-nonminimally-coupled-scalar field theories 
is marked by the presence of marginally-stable cloudy black-hole spacetimes (the term `scalar
clouds' is usually used in the physics literature \cite{Hodlit,Herlit} in order to describe spatially regular scalar fields 
which are linearly coupled to central supporting black holes). These critical spacetimes describe non-minimally coupled 
linearized scalar fields which, in the spherically symmetric case, 
are supported by central charged Reissner-Nordstr\"om black holes.

As nicely emphasized in \cite{Herch,Herr}, in order for the familiar (scalarless) charged black-hole solutions of general relativity 
(the Reissner-Nordstr\"om and Kerr-Newman black-hole spacetimes) to be valid solutions of the coupled 
Einstein-Maxwell-scalar field equations in the $\phi\to0$ limit, the 
scalar function $f(\phi)$, whose coupling to the Maxwell electromagnetic invariant of the charged spacetime can trigger the 
black-hole spontaneous scalarization phenomenon, should be characterized by the universal 
weak-field functional behavior $f(\phi)=1-\alpha\phi^2+O(\phi^4)$. 
The dimensionless physical parameter $\alpha$, which in previous studies has been assumed to be negative \cite{Herch,Herr,aaa1,aaa2,Hodsc1,Hodsc2,Hodnrx}, 
controls the strength of the direct scalar-field-electromagnetic-field non-trivial coupling. 

The intriguing spontaneous scalarization phenomenon of charged 
black holes in composed Einstein-Maxwell-nonminimally-coupled-scalar field theories 
is a direct consequence of the appearance of a spatially-dependent effective mass term, which in the weak-field $\phi\to0$ 
regime has the compact functional form $-\alpha{\cal F}/2$ \cite{Herch,Herr}, 
in the modified Klein-Gordon equation [see Eq. (\ref{Eq9}) below] of the non-trivially coupled scalar field. 
Interestingly, it has been established \cite{Herch,Herr,aaa1,aaa2,Hodsc1,Hodsc2,Hodnrx} that, 
for negative values of the dimensionless non-minimal coupling parameter $\alpha$, 
the effective mass term of the scalar field, which reflects its direct coupling to the Maxwell electromagnetic invariant of 
the spacetime, may become negative outside the horizon of the central supporting black hole. 

This important observation, which was first 
discussed in the context of negatively coupled ($\alpha<0$) scalar fields 
in the physically interesting works \cite{Herch,Herr}, implies that the composed 
black-hole-scalar-field effective potential term in the scalar Klein-Gordon wave equation 
may become {\it attractive} (negative) in the vicinity of the outer horizon, 
thus allowing the central charged black hole to support spatially regular bound-state configurations 
of the non-minimally coupled scalar fields. 

The main goal of the present paper is to reveal the physically interesting fact that massless scalar fields
with a non-minimal {\it positive} coupling ($\alpha>0$) to the Maxwell electromagnetic invariant ${\cal F}$ 
can also be supported in charged black-hole spacetimes. In particular, 
we shall explicitly prove that the positive-coupling black-hole spontaneous scalarization phenomenon in the 
composed Einstein-Maxwell-nonminimally-coupled-scalar field theory is {\it spin-and-charge} induced 
in the sense that only non-spherically symmetric Kerr-Newman black holes 
that possess {\it both} angular momentum and electric charge can support massless 
scalar fields with a non-minimal positive coupling to the Maxwell electromagnetic invariant 
of the spacetime. 

Below we shall explore the onset of the spontaneous scalarization phenomenon in spinning and charged Kerr-Newman 
black-hole spacetimes of the composed Einstein-Maxwell-nonminimally-coupled-scalar field theory with positive 
values of the physical parameter $\alpha$ \cite{Noteuniv}. Using analytical techniques, 
we shall derive a remarkably compact functional expression for the Kerr-Newman dimensionless 
rotation parameter $a/M$ which, for a given value of the black-hole dimensionless electric charge parameter $Q/M$ \cite{Noteam}, 
determines the critical boundary between bald (scalarless) Kerr-Newman black-hole spacetimes 
and hairy black-hole-scalar-field bound-state configurations. 

\section{Description of the system}

We shall study, using analytical techniques, the onset of the {\it positive}-coupling spontaneous scalarization 
phenomenon of spinning and charged Kerr-Newman black holes in the composed 
Einstein-Maxwell-nonminimally-coupled-massless-scalar field theory whose action 
is given by the integral expression \cite{Herch,Herr,aaa1,aaa2,Noteun}
\begin{equation}\label{Eq1}
S=\int
d^4x\sqrt{-g}\Big[R-2\nabla_{\alpha}\phi\nabla^{\alpha}\phi-f(\phi){\cal F}\Big]\  ,
\end{equation}
where the source term
\begin{equation}\label{Eq2}
{\cal F}=F_{\mu\nu}F^{\mu\nu}\
\end{equation}
is the Maxwell electromagnetic invariant of the spacetime. 
Below we shall explicitly prove that the presence of the 
direct scalar-electromagnetic non-trivial coupling term $f(\phi){\cal F}$ in the composed action (\ref{Eq1}) allows 
the existence of spontaneously scalarized black-hole spacetimes in the 
Einstein-Maxwell-nonminimally-coupled-massless-scalar field theory. 

The bald (scalarless) spinning and charged Kerr-Newman black-hole solution of the composed 
Einstein-Maxwell-nonminimally-coupled-scalar field theory (\ref{Eq1}) can be described, 
using the the Boyer-Lindquist spacetime coordinates $(t,r,\theta,\phi)$, 
by the curved line element \cite{ThWe,Chan}
\begin{eqnarray}\label{Eq3}
ds^2=-{{\Delta}\over{\rho^2}}(dt-a\sin^2\theta d\phi)^2+{{\rho^2}\over{\Delta}}dr^2+\rho^2
d\theta^2+{{\sin^2\theta}\over{\rho^2}}\big[a dt-(r^2+a^2)d\phi\big]^2\  ,
\end{eqnarray}
where the metric functions in (\ref{Eq3}) are given by the mathematically compact functional expressions
\begin{equation}\label{Eq4}
\Delta\equiv r^2-2Mr+a^2+Q^2\ 
\end{equation}
and
\begin{equation}\label{Eq5}
\rho^2\equiv r^2+a^2\cos^2\theta\  . 
\end{equation}
The conserved physical quantities $\{M,a,Q\}$ are respectively the mass, the angular momentum per unit mass, and the 
electric charge of the black hole. 
The horizon radii 
\begin{equation}\label{Eq6}
r_{\pm}=M\pm(M^2-a^2-Q^2)^{1/2}\
\end{equation}
of the Kerr-Newman black-hole spacetime (\ref{Eq3}) are determined by the roots of the metric function (\ref{Eq4}).

As emphasized above, it has been proved in \cite{Herch,Herr} that, 
in order for the familiar black-hole spacetimes of general relativity \cite{NoteGRKN} to be 
valid solutions of the composed field theory (\ref{Eq1}) in the $\phi\to0$ limit, 
the scalar coupling function $f(\phi)$ should be characterized by the weak-field universal functional 
behavior \cite{Herch,Herr}
\begin{equation}\label{Eq7}
f(\phi)=1-\alpha\phi^2+O(\phi^4)\  ,
\end{equation}
where the dimensionless physical parameter $\alpha$ controls the strength of the non-trivial interaction 
between the scalar field $\phi$ and the Maxwell electromagnetic invariant ${\cal F}$ of the spacetime. 
Intriguingly, one finds that the critical existence surface of the theory, 
which marks the boundary between bald (scalarless) black-hole spacetimes 
and hairy (scalarized) black-hole solutions of the composed Einstein-Maxwell-nonminimally-coupled-massless-scalar 
field theory (\ref{Eq1}), is universal for all non-linear scalar coupling functions that share the same 
weak-field leading-order expansion (\ref{Eq7}) \cite{Herch,Herr,aaa1,aaa2}.

The action (\ref{Eq1}), which characterizes the composed Einstein-Maxwell-nonminimally-coupled-massless-scalar 
field theory, yields the differential equation \cite{Herch,Herr,aaa1,aaa2}
\begin{equation}\label{Eq8}
\nabla^\nu\nabla_{\nu}\phi=\mu^2_{\text{eff}}\phi\
\end{equation}
for the non-minimally coupled scalar field, where 
the presence of the spatially-dependent effective mass term 
\begin{equation}\label{Eq9}
\mu^2_{\text{eff}}(r,\theta;M,Q,a)=-{1\over2}\alpha\cdot{\cal F}(r,\theta;M,Q,a)\ 
\end{equation}
in the Klein-Gordon equation (\ref{Eq8}) is a physically intriguing consequence 
of the non-trivial direct coupling between the scalar field $\phi$ and the Maxwell electromagnetic 
invariant ${\cal F}$ of the charged and spinning black-hole spacetime [see the action (\ref{Eq1})]. 

Hairy (spontaneously scalarized) black-hole solutions of the composed field theory (\ref{Eq1}) with negative values  
of the non-minimal coupling parameter $\alpha$ have been studied in \cite{Herch,Herr,aaa1,aaa2,Hodsc1,Hodsc2,Hodnrx}. 
In the present paper we shall reveal the physically intriguing fact that scalar fields 
with a non-minimal {\it positive} coupling ($\alpha>0$) to the Maxwell electromagnetic invariant (\ref{Eq2}) 
can also be supported in black-hole spacetimes. 

In particular, in the next section we shall prove that the spontaneous scalarization phenomenon 
of positively-coupled scalar-electromagnetic fields 
is a unique feature of black holes that possess a combination of both non-zero {\it spins} and 
non-zero electric {\it charges} (which, in principle, may take arbitrarily small non-zero values). 
Hence, we shall henceforth assume the characteristic relation 
\begin{equation}\label{Eq10}
a\cdot Q\neq 0\
\end{equation}
for the central supporting Kerr-Newman black holes. 

\section{Onset of positive-coupling spontaneous scalarization phenomenon 
in spinning and charged Kerr-Newman black-hole spacetimes}

In the present section we shall study the onset of the positive-coupling 
spontaneous scalarization phenomenon in the composed Einstein-Maxwell-nonminimally-coupled-massless-scalar 
field theory (\ref{Eq1}). In particular, using analytical techniques, we shall determine the critical 
onset-line $a_{\text{crit}}=a(Q)_{\text{crit}}$ of the composed physical system which, for a given non-zero value of the black-hole electric charge, 
determines the minimal value of the black-hole spin that can trigger the 
positive-coupling spontaneous scalarization phenomenon of the spinning and charged Kerr-Newman black holes. 

The presence of an attractive (negative) effective potential well in the modified Klein-Gordon wave equation (\ref{Eq8}) 
provides a necessary condition for the existence of non-minimally coupled scalar clouds (spatially regular bound-state field configurations) 
in the exterior region of the black-hole spacetime \cite{ChunHer,Hernn,Hodca,Hodjp}. 
Interestingly, from the functional expression (\ref{Eq2}) one deduces that, depending on the angular 
momentum $a$ of the spinning and charged black hole and the polar angle $\theta$, 
the spatially-dependent effective mass term (\ref{Eq9}), which characterizes the 
composed black-hole-nonminimally-coupled-scalar-field system, may become negative (thus representing an attractive potential well) 
in the vicinity of the horizon of the central Kerr-Newman black hole. 

In particular, the {\it onset} of the physically intriguing spontaneous scalarization phenomenon 
in the Kerr-Newman spacetime (\ref{Eq3}) is characterized by the critical functional relation \cite{ChunHer,Hernn,Hodca,Hodjp}
\begin{equation}\label{Eq11}
\text{min}\{\mu^2_{\text{eff}}(r,\theta;M,Q,a)\}\to 0^{-}\
\end{equation}
of the spatially-dependent effective mass term. 
For positive values of the coupling parameter $\alpha$ of the 
composed Einstein-Maxwell-nonminimally-coupled-scalar field theory (\ref{Eq1}), the characteristic critical 
relation (\ref{Eq11}) yields the functional behavior
\begin{equation}\label{Eq12}
\text{min}\{{\cal F}(r,\theta;M,Q,a)\}\to 0^{+}\
\end{equation}
for the Maxwell electromagnetic invariant of the spinning and charged spacetime 
at the onset of the spontaneous scalarization phenomenon. 

The non-vanishing components of the Maxwell electromagnetic tensor $F_{\mu\nu}$ in the 
axially-symmetric Kerr-Nemwan spacetime (\ref{Eq3}) 
are given by the (charge and spin dependent) expressions \cite{MXI1,MXI2}
\begin{equation}\label{Eq13}
F_{01}=-{{Q(r^2-a^2\cos^2\theta)}\over{(r^2+a^2\cos^2\theta)^2}}\ \ \ \ ; \ \ \ \ F_{13}=a\sin^2\theta\cdot F_{01}\
\end{equation}
and
\begin{equation}\label{Eq14}
F_{02}={{2Qa^2r\sin\theta\cos\theta}\over{(r^2+a^2\cos^2\theta)^2}}\ \ \ \ ; \ \ \ \ F_{23}={{r^2+a^2}\over{a}}\cdot F_{02}\  ,
\end{equation}
with the functional relation \cite{MXI1,MXI2}
\begin{equation}\label{Eq15}
{\cal F}=2\Big({{F^2_{02}}\over{a^2\sin^2\theta}}-F^2_{01}\Big)\  .
\end{equation}
Substituting the expressions (\ref{Eq13}) and (\ref{Eq14}) into (\ref{Eq15}), one finds 
the spatially-dependent functional expression \cite{MXI1,MXI2,NoteRN}
\begin{equation}\label{Eq16}
{\cal F}(r,\theta;M,Q,a)=-{{2Q^2(r^4+a^4\cos^4\theta-6r^2a^2\cos^2\theta)}\over{(r^2+a^2\cos^2\theta)^4}}\ 
\end{equation}
for the Maxwell electromagnetic invariant of the spinning and charged Kerr-Newman black-hole spacetime (\ref{Eq3}).

Taking cognizance of Eq. (\ref{Eq16}), one deduces the relations
\begin{equation}\label{Eq17}
\{{\cal F}(r,\theta;M,Q=0,a)\}=0\
\end{equation}
and
\begin{equation}\label{Eq18}
\{{\cal F}(r,\theta;M,Q\neq0,a=0)\}<0\
\end{equation}
for the spatially-dependent Maxwell electromagnetic invariant of 
the charged and spinning Kerr-Newman black-hole spacetime (\ref{Eq3}). 
These relations, together with the critical requirement (\ref{Eq12}) for the onset of the 
positive-coupling spontaneous scalarization phenomenon of Kerr-Newman black holes, imply that 
the spontaneous scalarization phenomenon of positively-coupled scalar-electromagnetic fields 
is a unique feature of black holes that have non-zero spins {\it and} non-zero electric charges [see the 
characteristic relation (\ref{Eq10})]. 

In particular, one deduces from Eqs. (\ref{Eq12}), (\ref{Eq17}), and (\ref{Eq18}) 
that non-spinning Reissner-Nordstr\"om black holes (with $a=0$ and ${\cal F}=-2Q^2/r^4<0$) 
and neutral Kerr black holes (with $Q=0$ and ${\cal F}\equiv 0$) 
{\it cannot} support spatially regular configurations of scalar fields which are positively coupled to the 
Maxwell electromagnetic invariant (\ref{Eq16}). 

We shall now prove that the critical onset-line $a_{\text{crit}}=a(Q)_{\text{crit}}$ of 
the composed Einstein-Maxwell-massless-scalar field theory (\ref{Eq1}), 
which describes cloudy bound-state Kerr-Newman-black-hole-nonminimally-coupled-linearized-massless-scalar-field 
configurations with the critical property (\ref{Eq12}), can be determined {\it analytically}. 
In particular, we shall explicitly show that the critical functional relation (\ref{Eq12}), 
which determines the onset of the positive-coupling spontaneous scalarization phenomenon of Kerr-Newman black holes, 
can be solved analytically. 

To this end, it proves useful to define the composed dimensionless variable
\begin{equation}\label{Eq19}
x\equiv {{a^2\cos^2\theta}\over{r^2}}\  ,
\end{equation}
in terms of which the Maxwell electromagnetic invariant (\ref{Eq16}) can be written 
in the remarkably compact dimensionless functional form
\begin{eqnarray}\label{Eq20}
{{r^4}\over{Q^2}}\cdot{\cal F}(x)=-{{2(1-6x+x^2)}\over{(1+x)^4}}\  .
\end{eqnarray}
Substituting the expression (\ref{Eq20}) into (\ref{Eq12}), one obtains the critical quadratic equation
\begin{equation}\label{Eq21}
1-6x+x^2=0\  ,
\end{equation}
which yields the critical value \cite{Notenpn}
\begin{equation}\label{Eq22}
x_{\text{crit}}=3-2\sqrt{2}\
\end{equation}
for the dimensionless ratio (\ref{Eq19}) at the onset 
of the positive-coupling spontaneous scalarization phenomenon of the spinning and charged 
Kerr-Newman black holes. 

Taking cognizance of Eq. (\ref{Eq19}) one deduces that, for a given {\it non}-zero value of the 
black-hole electric charge parameter $Q$, the minimally allowed value of the 
black-hole spin $a$ which is compatible with the positive-coupling spontaneous scalarization condition (\ref{Eq12}) can 
be inferred from the analytically derived dimensionless critical relation (\ref{Eq22}) with the 
maximally allowed value of the spatially dependent expression $\cos^2\theta/r^2$. 
In particular, the composed expression $\cos^2\theta/r^2$ is 
maximized by the coordinate values $(\cos^2\theta)_{\text{max}}=1$ with $r_{\text{min}}=r_+(M,a,Q)$ 
at the poles of the black-hole horizon, which yields 
\begin{equation}\label{Eq23}
\Big({{\cos^2\theta}\over{r^2}}\Big)_{\text{max}}={{1}\over{r^2_+}}\  .
\end{equation}

From Eqs. (\ref{Eq19}), (\ref{Eq22}), and (\ref{Eq23}) one finds the remarkably compact dimensionless critical relation
\begin{equation}\label{Eq24}
\Big({{a}\over{r_+}}\Big)_{\text{crit}}=\sqrt{2}-1\
\end{equation}
at the onset of the positive-coupling spontaneous scalarization phenomenon in the 
composed Einstein-Maxwell-nonminimally-coupled-scalar field theory (\ref{Eq1}). 
Interestingly, Kerr-Newman black holes with the property $a/r_+>(a/r_+)_{\text{crit}}$ are 
characterized by Maxwell electromagnetic invariants which are positive near the black-hole poles, 
in which case the effective scalar mass (\ref{Eq9}) in the modified Klein-Gordon wave equation (\ref{Eq8}) 
becomes negative in the near-horizon region, thus allowing the central Kerr-Newman black hole 
to support non-minimally coupled scalar field configurations.  

Taking cognizance of Eqs. (\ref{Eq6}) and (\ref{Eq24}), one obtains the functional expression 
\begin{equation}\label{Eq25}
\Big({{a}\over{M}}\Big)_{\text{critical}}={{1+\sqrt{1-2(2-\sqrt{2}){(Q/M)}^2}}\over{2\sqrt{2}}}
\ \ \ \ \text{for}\ \ \ \ 0<{{Q}\over{M}}\leq\sqrt{2\sqrt{2}-2}\
\end{equation}
for the critical existence-line of the 
composed Kerr-Newman-black-hole-nonminimally-coupled-massless-scalar-field configurations. 
The analytically derived functional expression (\ref{Eq25}) determines 
the minimally allowed value of the Kerr-Newman black-hole spin $a=a_{\text{critical}}(Q)$ that can trigger, 
for a given non-zero value of the black-hole electric charge parameter $Q$,
the positive-coupling spontaneous scalarization phenomenon in the 
composed Einstein-Maxwell-nonminimally-coupled-scalar field theory (\ref{Eq1}). 

Interestingly, one finds from (\ref{Eq25}) that the critical spin parameter of the cloudy Kerr-Newman black holes 
is a monotonically decreasing function of the black-hole electric charge parameter. 
In particular, the charge-dependent critical rotation parameter $a_{\text{critical}}=a_{\text{critical}}(Q)$, which determines 
the onset of the positive-coupling spontaneous scalarization phenomenon in the 
Einstein-Maxwell-nonminimally-coupled-scalar field theory (\ref{Eq1}), attains its global minimum value, 
\begin{equation}\label{Eq26}
\Big({{a}\over{M}}\Big)^*\equiv\min\Big\{\Big({{a}\over{M}}\Big)_{\text{critical}}\Big\}=\sqrt{2}-1\  ,
\end{equation}
at the extremal Kerr-Newman limit $(a^2+Q^2)/M^2\to1^-$ \cite{NoteKNE} with 
\begin{equation}\label{Eq27}
{{Q}\over{M}}=\sqrt{2\sqrt{2}-2}\  .
\end{equation}

It is physically interesting to point out that the classically allowed polar angular region for the positive-coupling near-horizon 
spontaneous scalarization phenomenon of spinning and charged Kerr-Newman black holes in the super-critical 
regime $a/r_+\geq(a/r_+)_{\text{crit}}$ is a monotonically increasing 
function of the dimensionless spin parameter $a/r_+$. 
In particular, the near-horizon Maxwell electromagnetic invariant (\ref{Eq16}) of the Kerr-Newman black-hole spacetime 
becomes positive in the polar angular range 
\begin{equation}\label{Eq28}
(3-2\sqrt{2})\cdot\Big({{r_+}\over{a}}\Big)^2\leq\cos^2\theta_{\text{scalar}}\leq1\  .
\end{equation}

The polar range (\ref{Eq28}), which can also be expressed in the remarkably compact form [see Eq. (\ref{Eq24})] 
\begin{equation}\label{Eq29}
\Big({{a_{\text{crit}}}\over{a}}\Big)^2\leq\cos^2\theta_{\text{scalar}}\leq1\  ,
\end{equation}
defines, in the super-critical regime $a\geq a_{\text{crit}}$, the classically allowed angular region for the spontaneous scalarization phenomenon of positively-coupled scalar 
fields in the near-horizon region of the spinning and charged Kerr-Newman black holes. 
In particular, the classically allowed angular region (\ref{Eq28}) is characterized by the limiting property 
\begin{equation}\label{Eq30}
(3-2\sqrt{2})\leq\cos^2\theta_{\text{scalar}}\leq1\
\end{equation}
for the maximally spinning Kerr-Newman black hole with $a/M\to 1^-$.  

\section{Summary and Discussion}

It has recently been proved \cite{Herch,Herr} (see also \cite{aaa1,aaa2,Hodsc1,Hodsc2,Hodnrx}) 
that charged black holes in composed Einstein-Maxwell-scalar field theories in which the scalar
field is non-trivially coupled to the Maxwell electromagnetic invariant of the charged spacetime with a {\it negative} coupling 
constant ($\alpha<0$) can support bound-state hairy configurations of the scalar field. 
Motivated by this physically intriguing observation, in the present paper we have revealed the fact that scalar 
fields which are {\it positively}-coupled ($\alpha>0$) to the Maxwell electromagnetic invariant can also be supported in 
asymptotically flat black-hole spacetimes. 

In particular, we have studied, using {\it analytical} techniques, 
the onset of the positive-coupling spontaneous scalarization phenomenon in spinning and charged Kerr-Newman 
black-hole spacetimes. The main results derived in this paper and their physical implications are as follows:

(1) We have revealed the physically intriguing fact that the 
black-hole spontaneous scalarization phenomenon of positively-coupled scalar-Maxwell fields is a unique feature 
of black holes that possess a non-zero combination ($a\cdot Q\neq0$) of angular momentum {\it and} electric charge [see Eq. (\ref{Eq10})]. 
Thus, the scalar-Maxwell positive-coupling spontaneous scalarization phenomenon of black holes is a {\it spin-charge} induced phenomenon. 

(2) We have proved that the critical black-hole dimensionless rotation parameter, 
\begin{equation}\label{Eq31}
{\hat a}\equiv {{a}\over{r_+}}\  ,
\end{equation}
which marks the boundary between the familiar (bald) Kerr-Newman black holes of the Einstein-Maxwell theory 
and hairy (scalarized) black holes in 
the composed Einstein-Maxwell-nonminimally-coupled-massless-scalar field theory (\ref{Eq1}) 
with a non-minimal positive scalar-field-Maxwell-electromagnetic-invariant coupling, 
is given by the remarkably compact analytically derived expression [see Eq. (\ref{Eq24})]
\begin{equation}\label{Eq32}
{\hat a}_{\text{crit}}=\sqrt{2}-1\  .
\end{equation}
Interestingly, spinning and charged Kerr-Newman black holes that lie above the critical onset-line (\ref{Eq32}) 
can support bound-state configurations of the positively-coupled scalar-Maxwell fields. 

The super-critical regime of the positive-coupling spontaneous scalarization phenomenon can be expressed in terms 
of the black-hole dimensionless physical parameters 
\begin{equation}\label{Eq33}
{\bar a}\equiv {{a}\over{M}}\ \ \ \ ; \ \ \ \ {\bar Q}\equiv {{Q}\over{M}}\
\end{equation}
in the simple form [see Eq. (\ref{Eq25})]
\begin{equation}\label{Eq34}
{\bar a}\geq{\bar a}_{\text{critical}}={{1+\sqrt{1-2(2-\sqrt{2}){\bar Q}^2}}\over{2\sqrt{2}}}
\ \ \ \ \text{for}\ \ \ \ 0<{\bar Q}\leq\sqrt{2\sqrt{2}-2}\  .
\end{equation}

(3) It has been proved that the minimal value of the dimensionless black-hole spin parameter ${\bar a}$ 
that can trigger the positive-coupling spin-charge induced 
spontaneous scalarization phenomenon in Kerr-Newman spacetimes is given by the 
compact expression [see Eqs. (\ref{Eq26}) and (\ref{Eq33})]
\begin{equation}\label{Eq35}
{\bar a}^*\equiv\min\{{\bar a}_{\text{critical}}({\bar Q})\}=\sqrt{2}-1\  .
\end{equation}

(4) It is worth emphasizing the fact that the positive-coupling spontaneous scalarization phenomenon in 
Kerr-Newman black-hole spacetimes can be triggered by arbitrarily small non-zero 
values of the black-hole electric charge parameter. 
In particular, from Eq. (\ref{Eq34}) one finds the critical black-hole spin 
\begin{equation}\label{Eq36}
{\bar a}_{\text{critical}}({\bar Q}\to 0)={{1}\over{\sqrt{2}}}\
\end{equation}
in the ${\bar Q}\to0$ limit of weakly-charged Kerr-Newman black holes. 

(5) We have proved that the classically allowed polar angular region for the positive-coupling near-horizon 
spontaneous scalarization phenomenon of black holes in the super-critical 
regime ${\hat a}\geq{\hat a}_{\text{crit}}$ is a monotonically increasing 
function of the Kerr-Newman dimensionless spin parameter ${\hat a}$. 
In particular, the classically allowed angular region for the black-hole spontaneous scalarization of positively-coupled 
scalar-Maxwell fields is characterized by the limiting near-extremal (${\bar a}\to 1^-$) behavior [see Eq. (\ref{Eq30})] \cite{NotesyKN}
\begin{equation}\label{Eq37}
0\leq\theta_{\text{scalar}}\lesssim 65.53^{\circ}\ \ \ \ \text{for}\ \ \ \ {\bar a}\to 1^-\  .
\end{equation}
This is the largest classically allowed polar angular region for the positive-coupling near-horizon 
spontaneous scalarization phenomenon of the spinning and charged Kerr-Newman black holes.

\bigskip
\noindent
{\bf ACKNOWLEDGMENTS}
\bigskip

This research is supported by the Carmel Science Foundation. I would
like to thank Yael Oren, Arbel M. Ongo, Ayelet B. Lata, and Alona B.
Tea for helpful discussions.



\begin{thebibliography}{99}

\bibitem{Bek1} J. D. Bekenstein, Phys. Rev. D {\bf 5}, 1239 (1972).

\bibitem{Her1} C. A. R. Herdeiro and E. Radu, Int. J. Mod. Phys. D {\bf 24}, 1542014
(2015).

\bibitem{Hodstationary} S. Hod, Phys. Lett. B {\bf 713}, 505 (2012);
S. Hod, Phys. Lett. B {\bf 718}, 1489 (2013) [arXiv:1304.6474]; S.
Hod, Phys. Rev. D {\bf 91}, 044047 (2015) [arXiv:1504.00009].

\bibitem{BekMay} A. E. Mayo and J. D. Bekenstein and, Phys. Rev. D {\bf 54}, 5059 (1996).

\bibitem{Hod1} S. Hod, Phys. Lett. B {\bf 771}, 521 (2017); S. Hod, Phys. Rev. D {\bf 96}, 124037 (2017).

\bibitem{ThWe} C. W. Misner, K. S. Thorne, and J. A. Wheeler, {\it Gravitation},
(W. H. Freeman, San Francisco, 1973).

\bibitem{Chan} S. Chandrasekhar, {\it The Mathematical Theory of Black
Holes}, (Oxford University Press, New York, 1983).

\bibitem{Herch} C. A. R. Herdeiro, E. Radu, N. Sanchis-Gual, and J. A. Font,
Phys. Rev. Lett. {\bf 121}, 101102 (2018).

\bibitem{Herr} P. G. S. Fernandes, C. A. R. Herdeiro, A. M. Pombo, E. Radu, and N.
Sanchis-Gual, Class. Quant. Grav. {\bf 36}, 134002 (2019)
[arXiv:1902.05079].

\bibitem{aaa1} D. C. Zou and Y. S. Myung, Phys. Rev. D {\bf 100}, 124055 (2019).

\bibitem{aaa2} P. G. S. Fernandes, Phys. Dark Univ. {\bf 30}, 100716 (2020) [arXiv:2003.01045] .

\bibitem{Hodsc1} S. Hod, Phys. Lett. B {\bf 798}, 135025 (2019) [arXiv:2002.01948].

\bibitem{Hodsc2} S. Hod, Phys. Rev. D {\bf 101}, 104025 (2020) [arXiv:2005.10268].

\bibitem{Hodnrx} S. Hod, The Euro. Phys. Jour. C {\bf 80}, 1150 (2020).

\bibitem{Hodlit} S. Hod, Phys. Rev. D {\bf 86}, 104026 (2012) [arXiv:1211.3202];
S. Hod, The Euro. Phys. Journal C {\bf 73}, 2378 (2013)
[arXiv:1311.5298]; S. Hod, Phys. Rev. D {\bf 90}, 024051 (2014)
[arXiv:1406.1179].

\bibitem{Herlit} C. A. R. Herdeiro and E. Radu, Phys. Rev. Lett. {\bf 112}, 221101 (2014).

\bibitem{Noteuniv} As nicely demonstrated in \cite{Herch,Herr}, the critical boundary between bald black-hole spacetimes 
and hairy (scalarized) black-hole spacetimes is universal for all 
Einstein-Maxwell-nonminimally-coupled-scalar field theories that share the same 
weak-field functional behavior $f(\phi)=1-\alpha\phi^2+O(\phi^4)$ of the non-minimal scalar coupling function.

\bibitem{Noteam} Here $M$, $J\equiv Ma$, and $Q$ are respectively 
the mass, angular momentum, and electric charge of the Kerr-Newman black hole. 
We shall assume, without loss 
of generality, the relations $a>0$ and $Q>0$ for the characteristic physical parameters of the 
spinning and charged Kerr-Newman black-hole spacetime. 

\bibitem{Noteun} We shall use natural units in which $G=c=\hbar=1$.

\bibitem{NoteGRKN} The most general black-hole solution of the Einstein-Maxwell field equations is described by 
the spinning and charged Kerr-Newman black-hole spacetime (\ref{Eq3}).

\bibitem{ChunHer} P. V. P. Cunha, C. A. R. Herdeiro, and E. Radu, Phys. Rev. Lett. {\bf 123}, 011101 (2019).

\bibitem{Hodca} S. Hod, Phys. Rev. D {\bf 102}, 084060 (2020) [arXiv:2006.09399].

\bibitem{Hernn} C. A. R. Herdeiro, A. M. Pombo, and E. Radu, Universe {\bf 7}, 483 (2021) [arXiv:2111.06442].

\bibitem{Hodjp} S. Hod, Jour. of High Energy Phys. {\bf 02}, 039 (2022) [arXiv:2201.03503].

\bibitem{MXI1} T. Adamo and E.T. Newman, The Kerr-Newman metric: A Review, arXiv:1410.6626 .

\bibitem{MXI2} I. Dymnikova and E. Galaktionov, 
Advan. in Math. Phys. {\bf 2017}, 1035381 (2017) [https://www.hindawi.com/journals/amp/2017/1035381/]; 
I. Dymnikova and E. Galaktionov, Universe {\bf 5}, 205 (2019).

\bibitem{NoteRN} Note that the (rather cumbersome) functional expression (\ref{Eq16}) reduces to the 
familiar compact expression ${\cal F}=-2Q^2/r^4$ for non-rotating Reissner-Nordstr\"om black holes \cite{Herch,Herr,aaa1,aaa2}.

\bibitem{Notenpn} Note that the second solution of the critical quadratic equation (\ref{Eq21}) 
is given by the dimensionless expression $x^{+}_{\text{crit}}=3+2\sqrt{2}>1$ and it therefore violates the physical requirement $x\leq1$ 
which follows from the inequalities $a/r\leq a/r_+\leq1$ [see Eq. (\ref{Eq6})] 
and $\cos^2\theta\leq1$. 

\bibitem{NoteKNE} Note that extremal Kerr-Newman black holes are characterized by the simple 
dimensionless relation $r_+/M=1$.  

\bibitem{NotesyKN} Note that, due to the reflection symmetry of the Kerr-Newman spacetime (\ref{Eq3}), 
the second classically allowed region for the positive-coupling near-horizon spontaneous scalarization phenomenon 
in the near-extremal ${\bar a}\to 1^-$ regime is characterized by the polar angular interval 
$114.47^{\circ}\lesssim\theta_{\text{scalar}}\leq180^{\circ}$. 

\end{thebibliography}
\end{document}